\documentclass[aps,twocolumn,showpacs]{revtex4}
\usepackage{graphicx}
\usepackage{dcolumn}
\usepackage{bm}

\begin{document}

\title{Transition from normal to ballistic diffusion in a one-dimensional impact system.}

\author{Andr\'e L.\ P.\ Livorati$^{1,2}$, Tiago Kroetz$^3$, Carl P. Dettmann$^{2}$, Iber\^e L. Caldas$^{4}$ and Edson D.\ Leonel$^{1}$}

\affiliation{$^1$ Departamento de F\'isica - UNESP - Universidade Estadual Paulista  Av. 24A 1515 - Bela Vista -
13506-900 - Rio Claro - SP - Brazil \\
$^2$ School of Mathematics - University of Bristol - Bristol - BS8 1TW - United Kingdom\\
$^3$  Departamento Acad\^emico de F\'isica - Universidade Tecnol\'{o}gica Federal do Paran\'{a}  UTFPR - Campus Pato Branco -  85503-390 - Pato Branco - PR - Brazil.\\
$^4$Instituto de F\'isica - IFUSP - Universidade de S\~ao
Paulo - USP  Rua do Mat\~ao  Tr.R 187 - Cidade Universit\'aria --
05314-970 -- S\~ao Paulo -- SP -- Brazil\\
\\ 
}

\begin{abstract}
We characterize a transition from normal to ballistic diffusion in a bouncing ball dynamics. The system is composed of a particle, or an ensemble of 
non-interacting particles, experiencing elastic collisions with a heavy and periodically 
moving wall under the influence of a constant gravitational field. The dynamics lead to a mixed phase space where chaotic orbits have a free path to 
move along the velocity axis, presenting a normal diffusion behavior. Depending on the control parameter, one can observe the presence of featured resonances, known as accelerator modes,
that lead to a ballistic growth 
of velocity. Through statistical and numerical analysis of the velocity of the particle, we are able to characterize a transition between the two regimes, where transport properties were used 
to characterize the scenario of the ballistic regime. Also, in an analysis of the probability of an orbit to
reach an accelerator mode as a function of the velocity, we observe a 
competition between the normal and ballistic transport in the mid range velocity.
\end{abstract}

\maketitle

\section{Introduction}
\label{sec1}
~~~~In 1949, Enrico Fermi \cite{fermi} proposed a mechanism to explain the origin of the high energies of the cosmic rays.
Fermi claimed that particles, which interacted with oscillating magnetic fields present in the cosmos, would on average exhibit
a gain of energy. This unlimited growth of energy was denoted Fermi acceleration (FA), and is mainly associated with normal diffusion in phase space,
where there is gain of kinetic energy \cite{litchenberg}. One may find in the literature examples of FA that may present transport
distinct from the normal diffusion, as exponential \cite{exp1,exp2,exp3,exp4}, ballistic \cite{kroetz15,kroetz16} or even
slower growths \cite{liv12,diego}. Also, interesting FA applications can be found in research areas such as plasma physics \cite{plasma1,plasma2}, 
astrophysics \cite{astrophys1,astrophys2}, atom-optics \cite{atom1,atom3}, and especially in billiard dynamics 
\cite{bil1,bil2,bil4,bil6,bil7}.\

The impact system under study in this paper is the so called bouncer (bouncing ball) model. Going back to Pustilnikov \cite{pustilnikov} the dynamics of the
system is composed of
a particle suffering elastic collisions with a vibrating platform under the influence of a constant gravitational field. 
The dynamics of the bouncing ball model has been studied for many years considering either non-dissipative and dissipative dynamics
\cite{everson,holmes,pieranski1,pieranski2,pieranski3}. For the non-dissipative version, the system basically behaves like the standard map 
in a local approximation \cite{litchenberg,liv12}, where some of the previous findings concerning the ballistic transport and accelerator 
modes (AM) in the standard map, serve as motivation background for this paper \cite{acel1,acel3,acel4,acel5,acel6,acel7,acel8}.
Yet, despite the simple dynamics, interesting applications for this system can be found in
dynamic stability in human performance \cite{tennis}, vibrations waves in a nanometric-sized mechanical contact system \cite{micros},
granular materials \cite{gran1}, experimental devices concerning normal coefficient of restitution \cite{gran3},
mechanical vibrations \cite{luo,barroso}, anomalous transport and diffusion \cite{klanges}, thermodynamics \cite{termo}, crisis between chaotic attractors \cite{crisis}, 
chaos control \cite{control1}, among others \cite{cd06,chaos08}.\

In the FA regime, the particle's velocity diffuses due to effectively random phases at which it reaches the platform. 
However, for certain values of the oscillation amplitude, it comes close to an attracting periodic orbit called an AM, in which the particle reaches the platform at the same phase, leading to linear growth of the velocity with the number of collisions.
Recent studies \cite{kroetz15,kroetz16} have considered the nature and focused on the localization (range of the control parameter) of the AM (ballistic modes), roughly described
as featured resonances in the phase space.\

In this paper we seek to understand the role of the accelerator modes in a transition from normal to ballistic diffusion in the dynamics of the bouncing ball model.
We focus in the description of the transport analysis and in the probability of an orbit to reach an accelerator mode (AM), emphasizing the transition from normal to ballistic diffusion. 
Through the analysis of the dispersion of the root mean square velocity, diffusion coefficient and the deviation of the mean square velocity by iteration,
we were able to characterize a diffusive transition in a range where a period-1 AM is active. Considering transport properties, such as the survival probability and escape rates for different velocity ranges, 
a description of the ballistic scenario for the AM was achieved. Also, an analysis of the probability of an orbit to reach the AM as a
function of the velocity leads us to observe a competition between normal and ballistic diffusion in a mid velocity range.
The results obtained in this work and the numerical procedures can be extended to other similar dynamical that present AM or superdiffusion 
in their dynamics.\

The paper is organized as follows: In Sec.\ref{sec2} we describe the details of the bouncing ball mapping and some chaotic properties. Section \ref{sec3} is devoted to the statistical 
analysis of the transition from normal to ballistic diffusion. In. Sec.\ref{sec4} we focus on the transport properties of the ballistic diffusive regime, where a competition between normal 
and ballistic diffusion was characterized.
Finally, in Sec.\ref{sec5} we draw some final remarks, conclusions and perspectives.\

\section{The model, the mapping and chaotic properties}
\label{sec2}
~~~~~This section is devoted to describing the impact system under study, the so called bouncing ball model, which consists of the motion of a 
particle that suffers elastic collisions with a heavy and periodically oscillating platform, under the presence of a constant gravitational field.
The dynamics of the system is described by a non-linear mapping \cite{liv12} for the variables velocity of the particle $v$ and time $t$
immediately after the $n^{th}$ collision of the particle with the moving wall. 

There are two distinct versions of the dynamical description: the complete one, which consists in considering the complete
movement of the time-dependent platform, and the simplified, that is often used to speed up numerical simulations, 
where the vibrating platform is set to be fixed, but the particle exchanges
momentum and energy with it, as if the platform were moving normally \cite{liv17}. Both approaches produce a very similar dynamics in both 
conservative and dissipative cases \cite{liv08,gabriel}. We consider in this paper the complete version, whose vibrating wall position
is given by $y_w(t_n)=\varepsilon\cos{w t_n}$, where $\varepsilon$ and $w$ are respectively the amplitude an
the frequency of oscillation.

Considering the flight time, which is the time that the particle goes up, stops with zero velocity, starts falling and
collides again with the vibrating wall, we define some dimensionless and more convenient variables: 
$V_n=v_n w/g$, $\epsilon=\varepsilon w^2/g$, where $V_n$ is the ``new dimensionless velocity", $g$ is the gravitational field and
$\epsilon$ can be understood as the ratio between the vibrating wall and the gravitational accelerations.
Also, measuring the time in terms of the number of oscillations of the vibrating wall, as $\phi_n=w t_n$, we finally end up with the following mapping 

\begin{equation}
T:\left\{\begin{array}{ll}
V_{n+1}=-({V_n^*}-{\phi_c})-2\epsilon\sin(\phi_{n+1})\\
\phi_{n+1}=[\phi_n+\Delta T_n]~~{\rm mod (2\pi)}\\
\end{array}
\right..
\label{eq1}
\end{equation}

The expressions for $V_n^*$, $\Delta T_n$, and the collision time defined as $\phi_c$ depend on what kind of collision
happens: (i) multiple collisions and; (ii) single collisions, where in both cases a 
transcendental equation is obtained for the condition that the position of the particle is 
the same as the position of the moving wall at the instant of the impact. For a more detailed description of Eq.(1), please check Refs.\cite{liv12,liv08}.

In the case of multiple collisions we have the scenario that after the particle enters in the
collision zone, $y_w(t_n)\in[-\epsilon,+\epsilon]$ and hits the moving platform, before
it leaves the collision zone, the particle suffers a second collision. It is also possible, depending
on the combination of $V_n$ and $\phi_n$, that the particle suffers many multiple
collisions \cite{metha}. In this case, the expressions for both $V_n^*$ and $\Delta T_n$ are given by $V_n^*=V_n$ and $\Delta T=\phi_c$.
The numerical value of $\phi_c$ is obtained as the smallest solution of an equation
$G(\phi_c)=0$ with $\phi_c\in(0,2\pi]$, where
\begin{equation}
G(\phi_c)=\epsilon\cos(\phi_n+\phi_c)-\epsilon\cos(\phi_n)-V_n\phi_c+{{
1}\over{2}}\phi_c^2~.
\label{eq2}
\end{equation}
If the function $G(\phi_c)$ does not have a root in the interval $\phi_c\in(0,2\pi]$, we can conclude that the particle
leaves the collision zone and a multiple collision no longer happens.

\begin{figure}[htb]
\begin{center}
\centerline{\includegraphics[width=8cm,height=11cm]{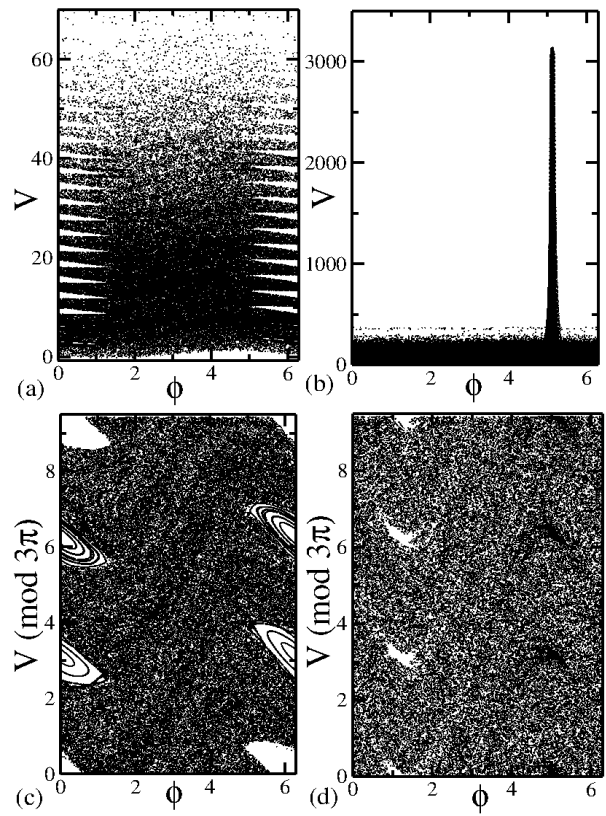}}
\end{center}
\caption{{\it Phase space for the complete dynamics of the bouncing ball model. In (a) and (c) $\epsilon=0.71$, in (b) and (d) $\epsilon=1.71$.}}
\label{fig1}
\end{figure}

The same discussion used for the function $G(\phi_c)$ also holds when we consider the case of single collisions. 
If the particle leaves the collision zone after a collision, goes up, reach null velocity (stops), and falls for an another collision we have
$V_n^*=-\sqrt{V_n^2+2\epsilon(\cos(\phi_n)-1)}$ and $\Delta
T_n=\phi_u+\phi_d+\phi_c$. Here $\phi_u=V_n$ denotes the time spent by the
particle in the upward direction up to reaching the null velocity,
$\phi_d=\sqrt{V_n^2+2\epsilon(\cos(\phi_n)-1)}$ corresponds to the time that
the particle spends from the place where it had zero velocity up to the
entrance of the collision zone at $\epsilon$.
Finally, $\phi_c$ is numerically obtained as the smallest solution of the equation $F(\phi_c)=0$ with $\phi_c\in[0,2\pi]$ where
\begin{equation}
F(\phi_c)=\epsilon\cos(\phi_n+\phi_u+\phi_d+\phi_c)-\epsilon-V_n^*
\phi_c+{{1}\over{2}}\phi_c^2~.
\label{eq3}
\end{equation}

The dynamics of the system undergoes some transitions as the control parameter $\epsilon$ changes \cite{litchenberg,liv12}, similar to the transitions found in the standard mapping.
For $\epsilon=0$, the system is integrable, and when $\epsilon$ is increased there is a transition from integrability to local chaos. In this range there is no FA, 
since the local chaotic sea is limited by invariant curves. If the control parameter goes beyond the critical one $\epsilon_c\approx0.2425$ \cite{liv12}, the system
faces a transition from local to global chaos. Such transition is crucial for the FA phenomenon to occur. Here, we have the destruction of the invariant spanning curves, allowing the union
of the local chaotic seas, so a chaotic orbit has a ``free path'' to diffuse along the velocity axis \cite{liv12}.

Figure \ref{fig1} shows the phase space for two different values of $\epsilon$, for $100$ different initial conditions iterated up to $10^3$ collisions.
The initial conditions were selected in an uniform distribution inside the range $V_0\in[\pi,2\pi]$ and $\phi_0\in[0,2\pi)$. 
In Fig.\ref{fig1}(a) we have $\epsilon=0.71$, and one can see a phase space with mixed properties and an increasing velocity, that is roughly uniformly distributed 
along the phase $(\phi)$ axis. One may also notice that the island structures repeat themselves according a $\pi$-size step \cite{kroetz15,kroetz16,liv12}. This repetition is more clear in 
Fig.\ref{fig1}(c), where the same phase space of Fig.\ref{fig1}(a) is plotted with the velocity axis mod $3\pi$.

Analyzing Fig.\ref{fig1}(b), where $\epsilon=1.71$, we can also see an increase in the velocity, but now there is a preferential phase, which dominates the dynamics, and the velocity 
reaches much higher values than Fig.\ref{fig1}(a). The behavior illustrated by Fig.\ref{fig1}(b) is the typical scenario of the influence of an accelerator mode (AM) in the dynamics, 
which causes a ballistic increase of the velocity. Figure \ref{fig1}(d) shows that this ballistic increase also obeys the repeating structure of the $\pi$-step size, where their position
are represented by the darker regions where $\phi\approx5$. Also, one could see in the anti-symmetric position of the AM some empty regions. Those are the decelerator modes (DM), which are 
unstable orbits in the sense that no typical initial condition can reach them, since they are repelling fixed points \cite{kroetz15}.

The difference between the two velocity regimes lies in the vibrating platform. For the regular FA (normal diffusion), the impacts sometimes occur when 
the platform is moving downwards, leading to an instantaneous loss of energy, but on average after several impacts a growth is observed.  
In contrast, for the AM (super diffusion) there is a periodic sequence of collisions with an overall gain in energy due to collisions when the platform is moving upwards.

Another interesting fact about the system dynamics, concerns the determinant of the Jacobian matrix, 
$det(J)={{V_n+\epsilon\sin(\phi_n)}\over{V_{n+1}+\epsilon\sin(\phi_{n+1})}}$ in the phase space. Since $det(J)$ can be greater or less than one, the map is not not symplectic in these 
coordinates \cite{kroetz15,liv12}. Note however that this gives us $$[V_{n+1}+\epsilon\sin(\phi_{n+1})]dV_{n+1}d\phi_{n+1}=[V_n+\epsilon\sin(\phi_n)]dV_nd\phi_n~,$$ which is equivalent to
$dE_{n+1}d\phi_{n+1}=dE_nd\phi_n$, in terms of the energy-like quantity $E_n=[V_n+\epsilon\sin(\phi_n)]^2/2$. For our impact system model, the existence of a set of variables in which the dynamics 
is area preserving is somewhat paradoxical, since this seems to rule out attracting periodic orbits such 
as accelerator modes. Similar behaviour regarding this non-symplectic properties can also be found in
the non-equilibrium Lorentz gas \cite{carl_lorentz1,carl_lorentz2}. The point is that for a translating (ballistic) periodic orbit, the periodicity of the system is
expressed in terms of variables $(V_n,\phi_n)$ in which the dynamics is not area preserving. Here, of course the periodicity is only approximate, improving as $V$ increases.

\begin{figure}[ht!]
\begin{center}
\centerline{\includegraphics[width=9cm,height=8cm]{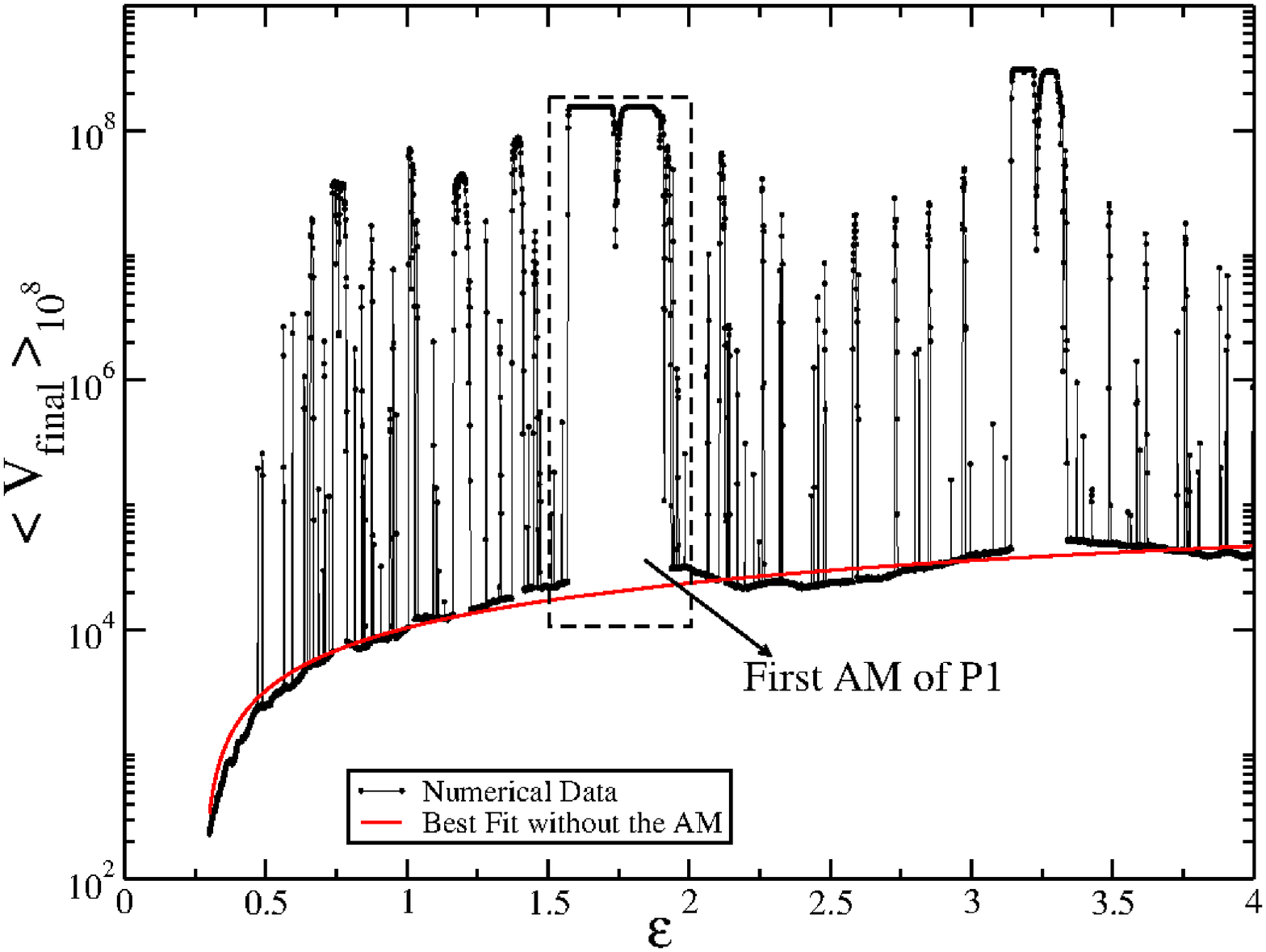}}
\end{center}
\caption{(Color online:){\it Final velocity after $10^8$ collisions as function of the control parameter $\epsilon$, where the several peaks denote the AM. The range of the first period-1 AM is depicted inside
the dashed box. Also, if we did not had any AM in the dynamics, a quadratic curve seems to fit well the dependence of the final velocity as function of $\epsilon$.}}
\label{fig2}
\end{figure}

To illustrate the contrast the AM plays in the dynamics, Fig.\ref{fig2} displays the behavior of an average over the value of the final velocity for an ensemble of $1000$ initial conditions,
at the end of $10^8$ iterations. One can see several distinguished peaks along the range of $\epsilon$, where each one of them represents an AM. Here, we show a range of interest in the dashed box 
including the first period-1 AM, which the stability is in a range of $\epsilon\in[\pi/2,\sqrt{1+\pi^2/4}]$ according to \cite{kroetz15,kroetz16,liv12}, 
and will be the range of $\epsilon$ in focus from now on. Also, if by any chance we could consider the dynamics without the AM influence, one could obtain a quadratic 
fit regarding the range of $\epsilon$ and the final velocity according to $V_{final}=676.88-9,741.8\epsilon+27,423\epsilon^2$.

\section{Transition from normal to ballistic diffusion}
\label{sec3}
~~~~~In this section we consider a statistical analysis for the dynamics of the bouncing ball model focusing in the transition from normal to ballistic diffusion.
We evaluate numerically and analytically the root mean square velocity, the diffusion coefficient and the dispersion of the mean squared velocity by collision, 
for a range of the control parameter $\epsilon$ where the period-1 AM is active.

Let us start by evaluating numerically the behavior of the root mean squared velocity, which is made by considering $V_{RMS}=\sqrt{\langle{V^2}\rangle}$, where
\begin{equation}
\langle{V^2}\rangle={1\over M} \sum_{i=1}^M {1\over n} \sum_{j=1}^n {(V_{i,j})}^2~,
\label{eq4}
\end{equation}
$M$ is the ensemble of initial conditions, and $n$ is the number of collisions (iterations). The average is taken along the orbit and along 
the ensemble of initial conditions. The initial conditions were chosen in the chaotic sea with velocity $V_0=\pi$ and the initial phase distributed uniformly in 
$\phi_0\in[0,2\pi)$. We took care to exclude any initial condition inside a stability island.

\begin{figure*}[ht!]
\begin{center}
\centerline{\includegraphics[width=17cm,height=12cm]{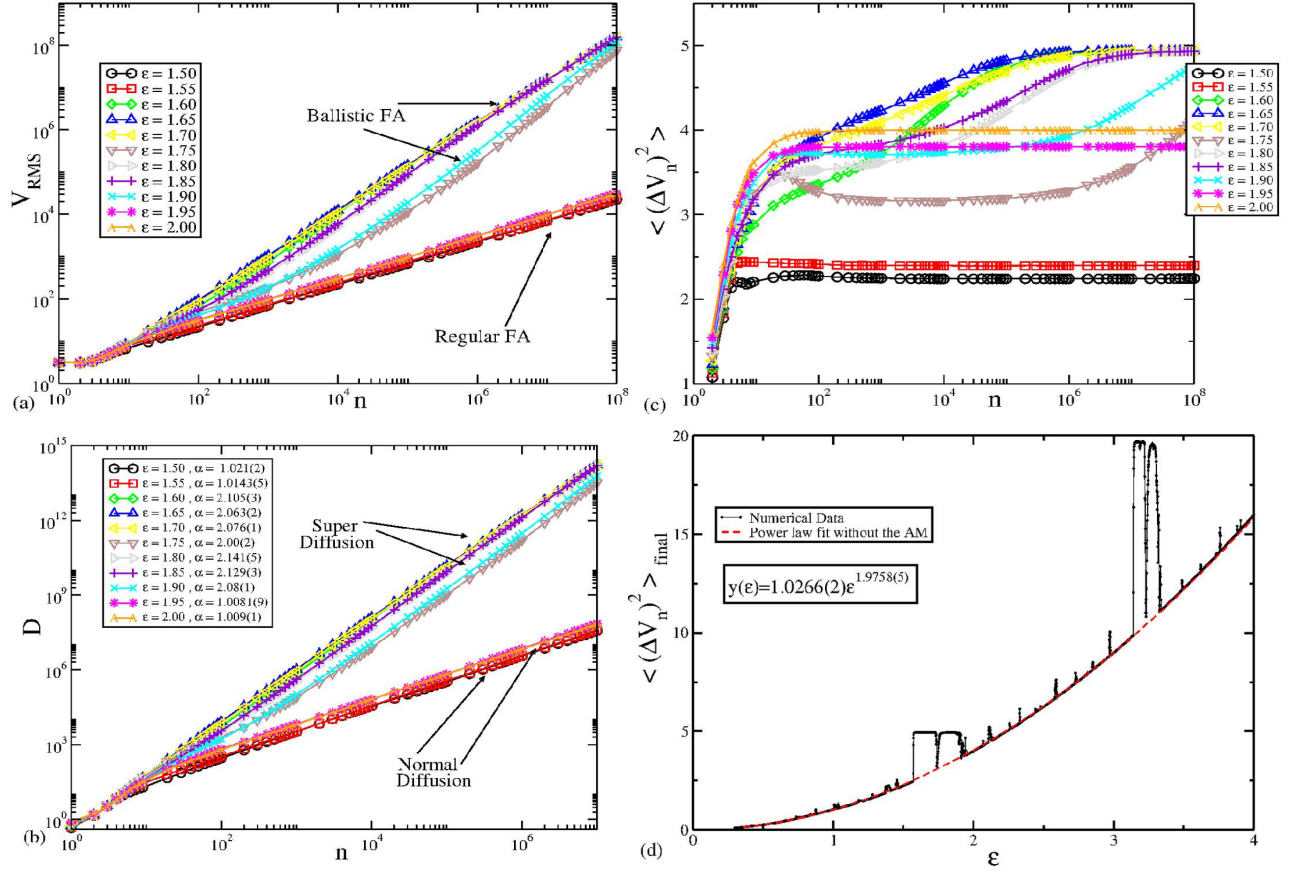}}
\end{center}
\caption{(Color online:){\it In (a)  the root mean square velocity for two regimes of velocity growth defined as regular Fermi acceleration and ballistic Fermi acceleration. In (b) the diffusion coefficient, given by Eq.(\ref{eq5a}) for the same values of the control parameter $\epsilon$ of (a)},
where the normal and ballistic diffusion behavior are depicted. In (c) we show the dispersion of the mean square velocity by collision iteration given by Eq.(\ref{eq6}), and in (d) we illustrate 
the final plateau of (c) as a function of $\epsilon$. }
\label{fig3}
\end{figure*}

One can see Fig.\ref{fig3}(a) that two distinct regimes of growth can be experienced by the dynamics. {\it(i)} the Regular Fermi Acceleration (RFA), and {\it(ii)} 
the Ballistic Fermi Acceleration BFA. In the RFA, the root mean square velocity curves grow according to $\sqrt{n}$, while in the BFA the $V_{RMS}$ curves obey a linear
growth, reaching higher velocities for very long times.

In order to obtain a contrast, we decided to compare the $V_{RMS}$ curves with the diffusion coefficient along the dynamics given by 
\begin{equation}
D=\lim_{n\to\infty}\frac{D_n}{2n}~,
\label{eq5a}
\end{equation}
where
\begin{equation}
D_n=\lim_{M\to\infty}\sum_{i=1}^M<(V_n^i-V_0^i)^2>~.
\label{eq5b}
\end{equation}

Here, $M$ is the size of the initial conditions ensemble and $n$ refers to the iteration number. We decided to stop the simulation at $n=10^7$ collisions
and considered $M=1000$ initial conditions, following the same line as the initial 
ensemble for the $V_{RMS}$ curves, since a higher value of $M$ would lead to similar results. 

After considering the numerical evaluation of the curves of $D~vs.~n$ in Fig.\ref{fig3}(b), we obtained by a power law fit,
a value of the exponent $\alpha$, expecting $D_n\sim n^\alpha$.
According to the literature, the $\alpha$ exponent defines what kind of diffusion we have in the system \cite{gaspard}. 
For $\alpha<1$, we have a sub diffusive regime, if $\alpha=1$ the normal diffusion (random walk) takes place, and finally if $\alpha>1$ we have the super 
diffusive regime.

Figure \ref{fig3}(b) shows the behavior of $D$ as function of the number of collisions $n$, for the same values of $\epsilon$ of Fig.\ref{fig3}(a).
One can see that for the control parameters where the curves present RFA, the diffusion coefficient has a linear behavior as $n$ evolves, with $\alpha\approx1$, which is in agreement with the normal diffusion theory.
On the other hand, for the control parameters that present BFA, the diffusion coefficient has a tendency to grow faster, with $\alpha\approx2$, indicating a ballistic diffusive regime in the dynamics.

One can see analyzing Figs.\ref{fig3}(a,b) a transition from normal diffusion to ballistic diffusion in the dynamics, when the first period-1 AM acquires stability. To illustrate such transition, let us study another variable of interest, 
which is the dispersion of the mean square velocity \cite{gaspard} by collision iteration, that is given by
\begin{equation}
\langle{(\Delta V_n})^2\rangle=\lim_{M\rightarrow\infty} {1\over M}\sum_{i=1}^{M}(V_{n+1}^i-{V_n}^i)^2~,
\label{eq6}
\end{equation}
where again $M$ is the same ensemble of initial conditions, the index $i$ denotes the $M$ particles and $V_n$ is the velocity after $n$ iteration 
of the i$^{th}$ particle.

The expressions hold in Eqs.(\ref{eq5b}) and (\ref{eq6}) look like the same, but they differ in the way the averages are evaluated. In equation (\ref{eq5b}) we consider the average over the initial condition
$V_0$, while in Eq.(\ref{eq6}) the average is taken over the difference between the velocities $V_n$ and $V_{n+1}$ at each iteration.

Figure \ref{fig3}(c) shows the behavior of $\langle{(\Delta V_n})^2\rangle$ as function of the number of collisions $n$, for the same values of $\epsilon$ of Figs.\ref{fig3}(a,b).
One can see that for the control parameters where the curves presents RFA, the curves establish themselves in a constant plateau after a
few iterations. On the other hand, for the control parameters that present BFA, the curves have a tendency of growth for short and medium times, 
and then bend towards a higher constant plateau for very long times. In particular, we can see a transition from normal to ballistic diffusion when $\epsilon=1.75$ and
$\epsilon=1.90$, where the plateaus are nearly constant until $10^6$ iterations, and then they bend towards a growth regime to the same region where the other  
curves of $\langle{(\Delta V_n})^2\rangle$ converged when the BFA is active.

As an attempt to explain the convergence plateaus and the transitions from normal to ballistic diffusion, let us made an analytical analysis of the
statistical properties of the velocity. One can consider the recurrence expression for the velocity of the mapping (\ref{eq1}), and take the square of both sides of it, obtaining then 
${V_{n+1}}^2={V_n^*}^2-2V_n\phi_c+{\phi_c}^2-4\epsilon\sin(\phi_{n+1})(-V_n^*+\phi_c)~\\
+4\epsilon^2\sin^2(\phi_{n+1})$. Since when the AM is active, we have only the case of single collisions with the moving platform \cite{kroetz15}, so the term $\phi_c$ of the mapping (\ref{eq1}) should be
obtained from $F(\phi_c)$ on Eq.(\ref{eq3}), where $\phi_c=V_n^*\pm\sqrt{{V_n^*}^2-2\epsilon[\cos(\phi_{n+1})-1]}$ is obtained from solving a quadratic equation and 
should be replaced in the above expression. After straightforward algebra and evaluating an average 
over all terms in the interval $\phi\in[0,2\pi]$, where for the terms depending of the phase, 
we have zero for $\sin(\phi_{n+1})$ and $\sin(\phi_n)$, and $1/2$ for $\sin^2(\phi_{n+1})$, we finally end up with
\begin{equation}
\langle(\Delta V)^2\rangle=2\epsilon^2~,
\label{eq7}
\end{equation}
where $\langle(\Delta V)^2\rangle=\overline{(V_{n+1})^2}-\overline{(V_n)^2}$.

Combining the results obtained in Eq.(\ref{eq6}), with the expression hold in Eq.(\ref{eq7}), we achieve that in the normal diffusion regime we have $\langle{(\Delta V_n})^2\rangle\propto\epsilon^2$.
Figure \ref{fig3}(d) shows the behavior of the final plateau established by $\langle{(\Delta V_n})^2\rangle$ according Eq.(\ref{eq6}) as function of an extensive range of the control parameter $\epsilon$. One can see that
as far $\epsilon$ grows, the value of the dispersion of the mean square velocity also grows. If we consider the evolution of $\langle{(\Delta V_n})^2\rangle~vs.~\epsilon$ disregarding the AM, we obtain a power law 
fitting according to $\langle{(\Delta V_n})^2\rangle=y(\epsilon)=1.0266(1)\epsilon^{1.9758(5)}$, which is very close to the expected theoretical result where $\langle{(\Delta V_n})^2\rangle\propto\epsilon^2$. 


\section{Transport and Survival probability}
\label{sec4}
In this section we address the transport of orbits for the range of $\epsilon$ when the AM is active, i. e., when the system is under a ballistic diffusive regime of dynamics.
A natural observable allowing the study of the statistical properties of the transport, in particular $\rho(n)$, the probability (given a suitable distribution
of initial conditions) that an orbit does not escape through a hole until a time $n$. Here, the hole is defined as a predefined subset of the phase space.
The most important aspect of this analysis is that the escape rate is very sensitive to the system dynamics \cite{meiss,alltman}. For strongly chaotic systems the decay is typically
exponential \cite{doubling,liv_fum,mexico}, while systems that present mixed phase space the decay can be slower, presenting a mix of exponential 
with a power law \cite{liv12}, or even stretched exponential decay \cite{stretched}. 

Since for our modeling the FA phenomenon is inherent in the system dynamics, we consider that an initial condition had escaped through the hole
if its velocity is equal, or higher than $V=V_{hole}$. Then, we save in a vector the iteration in which the orbit had escaped, and then we build a frequency 
histogram for the escape, according the escape iteration. Here, the hole is set as a 'line' in the velocity axis, with arbitrary phase.

The survival probability, described in terms of escape formalism \cite{gaspard,meiss,alltman}, is then obtained by the integration of this 
escape frequency histogram, as
\begin{equation}
\rho(v,n)={1\over M} {\sum_{j=1}^M} N_{rec}(j)~,
\label{eq11} 
\end{equation}
where the summation is taken along an ensemble of $M=10^6$ initial conditions chosen along the chaotic sea. Here, $v$ is set as the escape velocity, or the hole position in the velocity axis.
The term $N_{rec}(j)$, denotes the number of initial conditions that did not escape through the holes until the j-th collision \cite{gaspard}.
The initial conditions were set as: the initial velocity was always the same as $V_0=\pi$, and the initial phase $\phi_0$ was distributed along $\phi_0\in[0,2\pi)$, where we took 
an extra care to not chose any initial phase that could belong to a stability island, or otherwise the statistics would be damaged.

\begin{figure}[ht!]
\begin{center}
\centerline{\includegraphics[width=8cm,height=12cm]{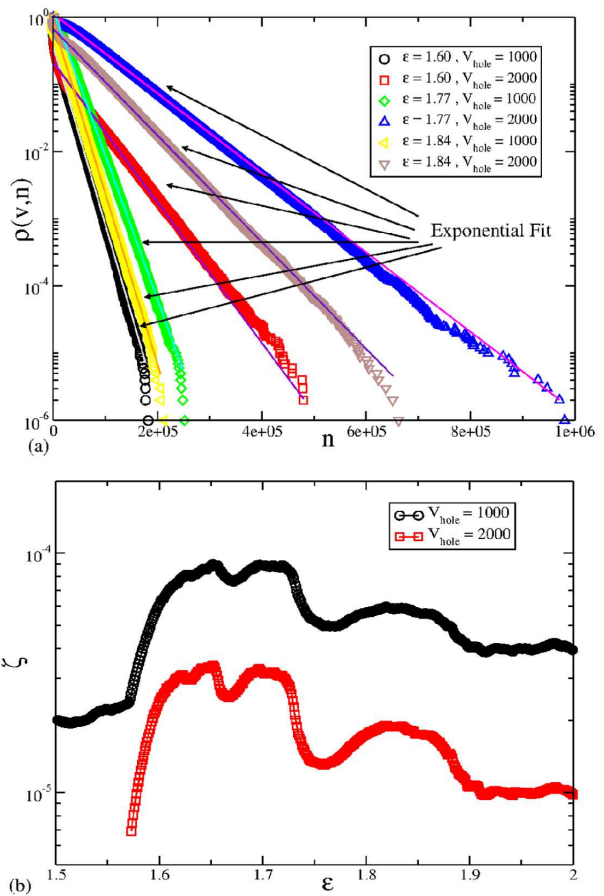}}
\end{center}
\caption{(Color online:){\it In (a) we have the behavior of $\rho(v,n)$ for a range where the AM is active for two different holes, where all the decays are exponential. In (b) we have 
the $\zeta$ exponent for both holes as function of $\epsilon$. }}
\label{fig4}
\end{figure}

At first, we selected two escape velocities as $V_{hole}=1000$ and $V_{hole}=2000$, and evaluated the dynamics for the range of the control parameter $\epsilon$ where the AM of period-1 is active.
Figure \ref{fig4}(a) shows the behavior of $\rho(v,n)$ for a few values of $\epsilon$ for both holes. One can see basically an exponential decay as 
\begin{equation}
\rho(v,n)=A\exp(-\zeta n)~,
\label{eq12}
\end{equation}
where the value of $\zeta$ may depend on $\epsilon$ and the selected hole. Here the iterations were evaluated up to $10^6$ collisions.

In Fig.\ref{fig4}(b) we display the behavior of $\zeta$ for two values of the escape velocity hole considering the whole range of $\epsilon\in[1.5,2.0]$, which includes the first AM of period-1.
One can see that the peaks scenario for both holes is quite similar to the range of variation of Fig.\ref{fig2}, indicating in which range of the control parameter $\epsilon$ the AM has 
more influence in the dynamics. 

\begin{figure}[ht!]
\begin{center}
\centerline{\includegraphics[width=8cm,height=10cm]{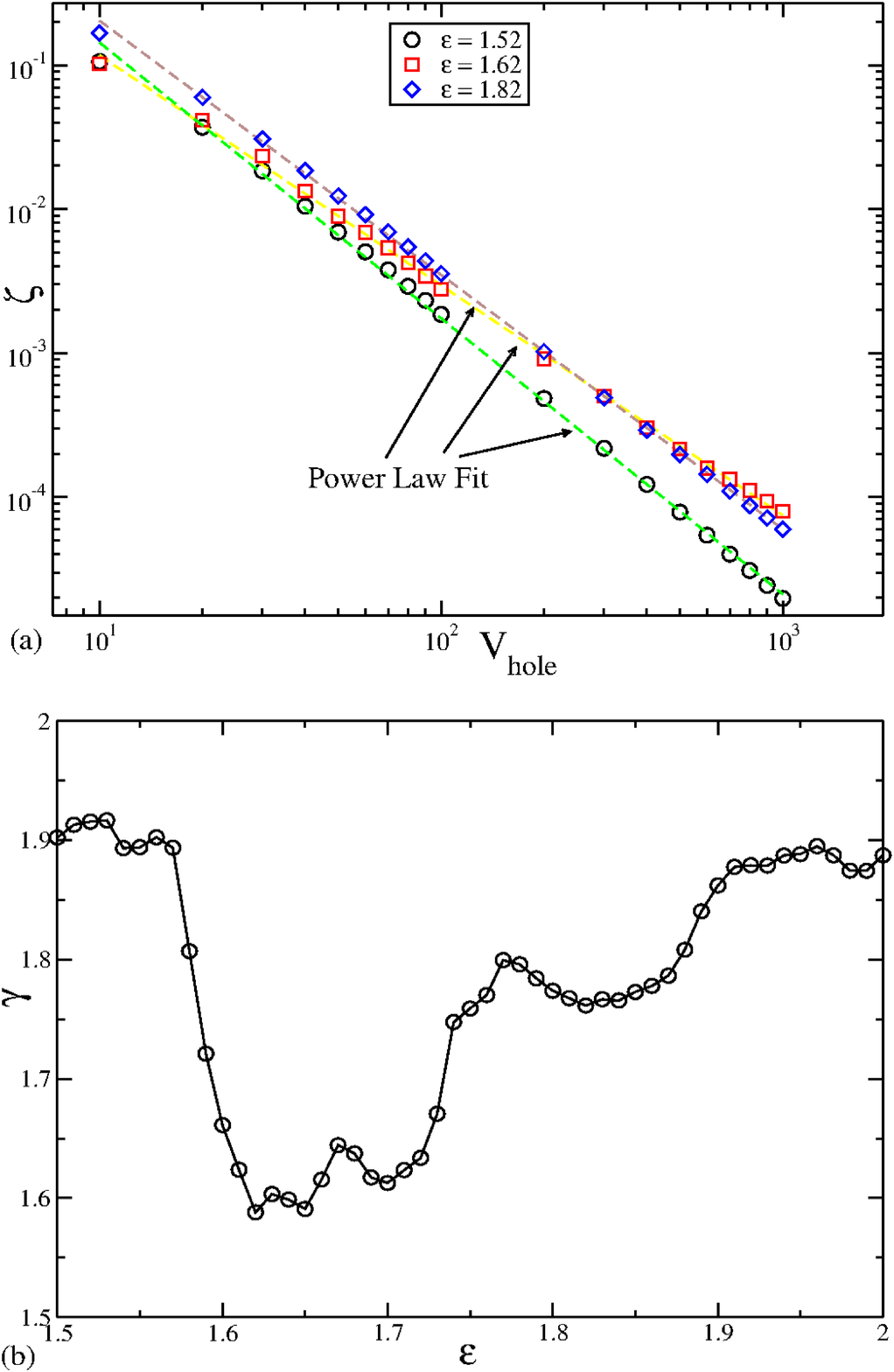}}
\end{center}
\caption{(Color online:){\it In (a) escape rate $\zeta$ for the $20$ different velocity holes for some values of $\epsilon$, where a power law seems to be the best fit. In (b) the power law exponent $\gamma$
obtained in (a) as function of $\epsilon$.}}
\label{fig5}
\end{figure}

\begin{figure}[ht!]
\begin{center}
\centerline{\includegraphics[width=8cm,height=8cm]{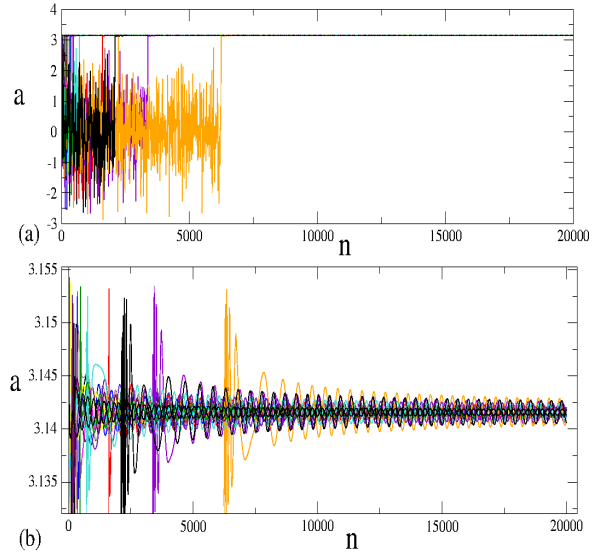}}
\end{center}
\caption{(Color online:){\it In (a) the linear coefficient $a$ as function of $n$, and in (b) a zoom-in in the convergence to $\pi$ region.}}
\label{fig6}
\end{figure}

In order to understand better the influence of the escape velocity, we have selected now $20$ different holes, equally split among two decades
between $V_{hole}\in[10,1000]$ in the range of $\epsilon$ where the first AM of period-1 is active. For all holes for the whole range of $\epsilon$ we observed 
an exponential decay rate of 
$\rho(v,n)$, just like the ones observed in Fig.\ref{fig4}(a). So, in Fig.\ref{fig5}(a) we show the behavior of every escape rate $\zeta$ for the $20$ different velocity holes, for 
the same values of $\epsilon$, where a power law of the type 
\begin{equation}
\zeta\propto (V_{hole})^\gamma~,\
\label{eq13} 
\end{equation}
is the best fitting in the numerical data.

Figure \ref{fig5}(b) shows the behavior of the $\gamma$ exponent as $\epsilon$ is ranged. One can see there is a slight decay in the value of $\gamma$ in the
range of $\epsilon$ when the AM is active. Also, the plot of $\gamma~vs.~\epsilon$ is similar in a upside down manner to Fig.\ref{fig4}(b), indicating where the AM is stronger 
and weaker.

The results provided in Figs.\ref{fig4} and \ref{fig5} are in good agreement with the results obtained in \cite{kroetz15,kroetz16}, where an analysis of the nature and the bifurcation process of the first period-1 AM 
was made. The peaks represent the regions where the AM is stronger and more active, while  the valleys are related with bifurcations and suddenly lost of stability. An example is the period-3 
catastrophe \cite{acel3,acel4}, for $\epsilon \approx 1.75$. However, there are some issues that one could argue about the AM dynamics. For instance, how do we know if an orbit reached the AM? What about the dependence 
on the velocity? These questions will be the focus of the paper from now on.

Let us start by defining a criterion for the convergence to the AM. Since we are interest in the range of $\epsilon$ where the first period-1 AM is active, 
we already know from \cite{kroetz15,kroetz16} and from Fig.\ref{fig1}(b,d) there is a step-size of $\pi$ for the AM, and from Fig.\ref{fig3}(a), we know there is a linear growth of the $V_{RMS}$. So, a linear regression of the type 
$V_n=an+b$ should provide us $a\approx\pi$. In order to consider this criterion, we evaluated a linear regression in the dynamical evolution at every $30$ steps of $n$, with a tolerance of $\delta=\pm0.001$, in order
to have a better statistics for our analysis.
Figure \ref{fig6}(a) displays the behavior of $a$ as $n$ evolves, for a few orbits considering $\epsilon=1.61$, and in Fig.\ref{fig6}(b) there is a zoom-in window 
for $a\approx\pi$. Both figures show for long times the linear coefficient converging to $\pi$. So, using this convergence scenario seems a good criterion to establish if an orbit reached the AM of period-1.

\begin{figure}[ht!]
\begin{center}
\centerline{\includegraphics[width=9cm,height=13cm]{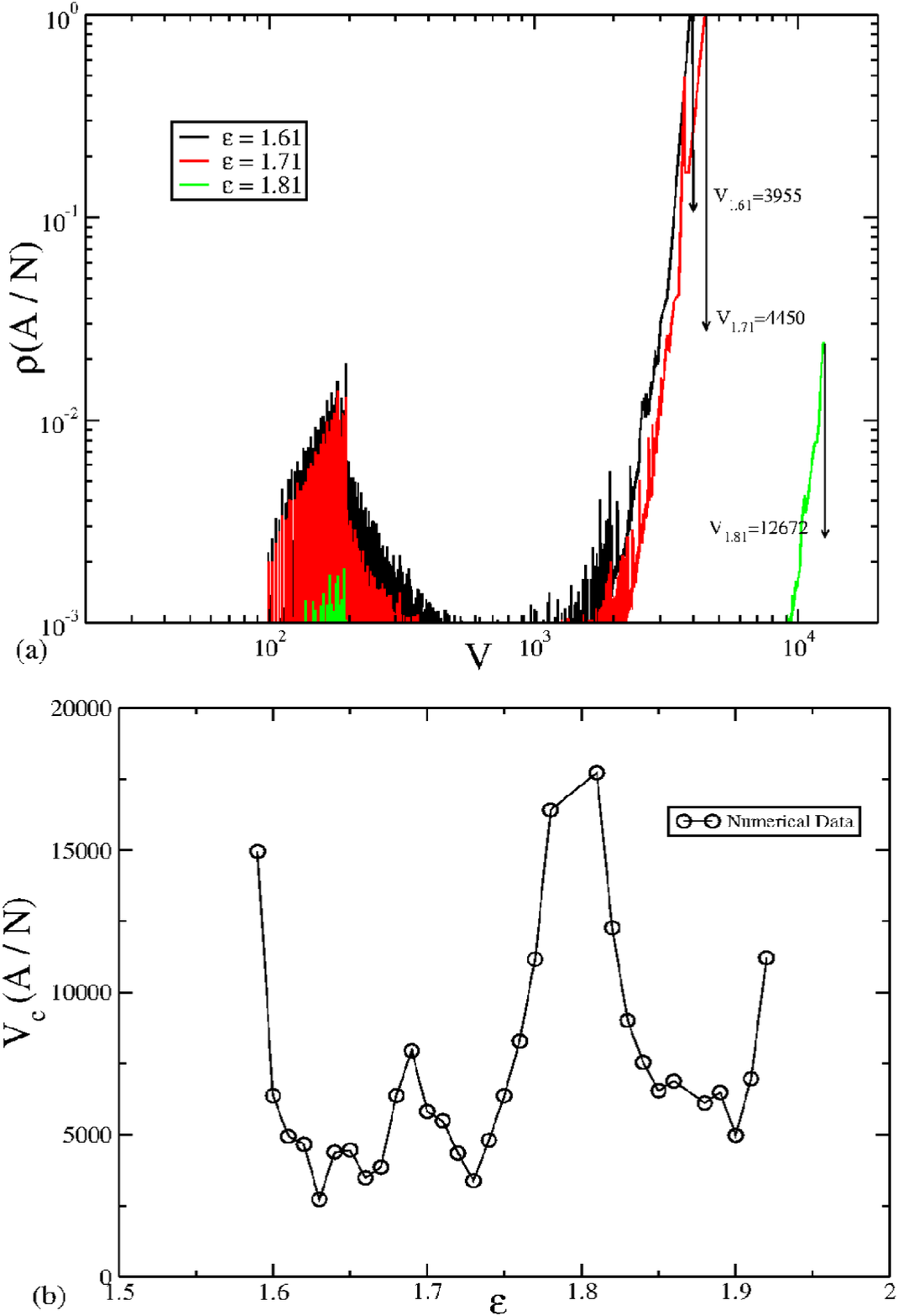}}
\end{center}
\caption{(Color online:){\it In (a) the probability of an ensemble of initial conditions to reach the AM as function of the velocity for some values of $\epsilon$. The first Gaussian-like peak represents a competition between normal
and ballistic diffusion, and the maximum of $\rho(A/N)$ denotes the critical velocity for which the AM was finally achieved. 
In (b) the critical velocity $V_c$, obtained in (a), as function of $\epsilon$. 
The similar peaks scenario with Fig.\ref{fig5}(b) gives robustness to our probability analysis.}}
\label{fig7}
\end{figure}

Moving forward, we are now interested in investigate the dependence on the velocity for an orbit that reached an AM. Using the linear regression convergence criterion, we created histograms of 
frequencies as a function of the velocity of the orbit for two different dynamical cases: {\it(i)} before reaching the AM, which we label as $N$ (for normal diffusion), and {\it(ii)} when the orbit is at the AM, 
which we label as $A$ (for accelerator mode). Each labeled vector has a range from $V\in[-\epsilon,20000]$, and this range was split in $10^5$ equal parts (boxes). 

For each initial condition starting with low velocity, we do the following procedure: At each collision, we keep adding unity to the relevant $N$ box, until the linear coefficient reaches the value of $a=\pi\pm0.001$.
After that, we know the orbit reached the AM, then we add one to the $A$ box, stop the simulation and start a new initial condition. The addition to the relevant $N$ or $A$ box was made considering the convergence
criterion at every $30$ collisions.

Figure \ref{fig7}(a) shows the behavior of the probability $\rho(A/N)$, which is the ratio between the histograms for Accelerated (A) and Normal (N) dynamics, already normalized according the ensemble of
initial conditions versus velocity for some values of $\epsilon$. Here we can depict two distinct regimes. 

The first one is when
the velocity is in a range about $V\in[100,500]$. One can see a peak in a Gaussian-like shape, that does not seems to depend of $\epsilon$ 
(at least in the velocity range). We believe in this range, there is a 
competition between the normal diffusion $N$ and the ballistic diffusion $A$, where some orbits may achieve the AM really fast and others can take longer times. 

The other scenario concerns the maximum of $\rho(A/N)$, which varies with $\epsilon$.  As far as we understand when the maximum is reached, 
the respective velocity can be considered a critical one, where at this velocity we may know that all orbits reached the AM. 
In particular, for $\epsilon=1.81$, where the maximum of $\rho(A/N)$ does not reach unity, a possible explanation is that for this value of $\epsilon$,
the AM is not so influential, as one can observe
in the peaks of Fig.\ref{fig3}(d). Another possibility is that some of the orbits might get trapped in a stickiness regime, and this anomalous behavior would damage the statistics.

Considering now the value of the critical velocity, where $\rho(A/N)$ reaches its maximum, in Fig.\ref{fig7}(b) we can observe the same peaks scenario as observed in Fig.\ref{fig5}(b), 
as we range the control parameter $\epsilon$, when the AM is active. This result gives robustness to our analysis of probability as function of the velocity.

\section{Final Remarks and Conclusions} 
\label{sec5}

To summarize, we have investigated the dynamics of a particle undergoing elastic collisions under the influence of a constant gravitational field in a domain composed 
by a heavy and periodic moving platform.
A nonlinear mapping was obtained and a mixed phase space was characterized composed by a chaotic sea and KAM islands, where the particle has a free path to diffuse in the velocity, leading the
dynamics to exhibit unlimited growth of energy (velocity), known as Fermi Acceleration.

Depending on the control parameter one may observe regular and/or ballistic FA, where the RFA is originated by normal diffusion in the chaotic sea, while the BFA is due the presence of accelerator modes in the dynamics, 
leading to ballistic behavior. We characterized a transition from normal to ballistic diffusion while the first period-1 AM is active. Statistical and numerical analysis for the root
mean square velocity, the diffusion coefficient and the deviation of the mean square velocity by iteration were evaluated. Also, a remarkable analytical agreement
was achieved regarding a dependence of the square of the control parameter. 

Considering transport properties, such as the survival probability and escape rates for different velocity ranges, a description of the ballistic scenario 
for the AM was made, where we found that some ranges
of control parameters are more influential than others, since the first AM of period-1 undergoes a series of bifurcations and loss of stability during this 
particular range. Also, an analysis of the probability of an orbit
to reach the AM as a function of the velocity leads us to interpret a competition between normal and ballistic diffusion in a mid velocity range. As a next step, we 
intend to investigate how different and higher periods of the AM influence the transport properties and the transition from normal to ballistic diffusion from local
and global points of view.

\acknowledgments
ALPL acknowledges FAPESP (2014/25316-3) and FAPESP (2015/26699-6) for financial support. CPD thanks EPSRC grant (EP/N002458/1).
ILC thanks FAPESP (2011/19296-1) and CNPq, EDL thanks FAPESP (2017/14414-2) and CNPq (303707/2015-1). 
ALPL also thanks the University of Bristol for the kindly hospitality during his stay in UK. This research was
supported by resources supplied by the Center for Scientific Computing (NCC/GridUNESP) of the S\~ao Paulo State University (UNESP).

\end{document}